\documentclass[10pt,twocolumn,letterpaper]{article}

\usepackage[pagenumbers]{wacv} 

\definecolor{wacvblue}{rgb}{0.21,0.49,0.74}
\usepackage[pagebackref,breaklinks,colorlinks,allcolors=wacvblue]{hyperref}
\usepackage{multirow}
\usepackage{float}
\usepackage{longtable}
\usepackage{makecell}

\title{A Graph-Based Framework for Interpretable\\ Whole Slide Image Analysis}

\usepackage{setspace}
\author{
\begin{tabular}{ccc}
Alexander Weers$^{1,3}$ & Alexander H. Berger$^{1,6}$ & Laurin Lux$^{1,3,6}$ \\
Peter Schüffler$^{3,5}$ & Daniel Rueckert$^{1,2,3,4}$ & Johannes C. Paetzold$^{2,6}$
\end{tabular}\\
\small
$^{1}$School of Computation, Information and Technology, Technical University of Munich, DE \\
\small
$^{2}$Department of Computing, Imperial College London, UK \\
\small
$^{3}$Munich Center for Machine Learning, DE \\
\small
$^{4}$Munich Data Science Institute, Technical University of Munich, DE \\
\small
$^{5}$Institute of Pathology, TUM School of Medicine and Health, Technical University of Munich, Germany \\
\small
$^{6}$Weill Cornell Medicine, Cornell University, New York City, NY, USA\\
    {\tt\small \href{mailto:alexander.weers@tum.de}{alexander.weers@tum.de}; \href{mailto:jpaetzold@med.cornell.edu}{jpaetzold@med.cornell.edu}}
}

\usepackage{xtab}

\begin{document}
\maketitle

\begin{abstract}
The histopathological analysis of whole-slide images (WSIs) is fundamental to cancer diagnosis but is a time-consuming and expert-driven process.
While deep learning methods show promising results, dominant patch-based methods artificially fragment tissue, ignore biological boundaries, and produce black-box predictions.
We overcome these limitations with a novel framework that transforms gigapixel WSIs into biologically-informed graph representations and is interpretable by design.

Our approach builds graph nodes from tissue regions that respect natural structures, not arbitrary grids.
We introduce an adaptive graph coarsening technique, guided by learned embeddings, to efficiently merge homogeneous regions while preserving diagnostically critical details in heterogeneous areas.
Each node is enriched with a compact, interpretable feature set capturing clinically-motivated priors.
A graph attention network then performs diagnosis on this compact representation.

We demonstrate strong performance on challenging cancer staging and survival prediction tasks.
Crucially, our resource-efficient model ($>$13x fewer parameters and $>$300x less data) achieves results competitive with a massive foundation model, while offering full interpretability through feature attribution.
Our code is publicly available at \url{https://anonymous.4open.science/r/pix2pathology-5680}.

\end{abstract}

\section{Introduction}
Histopathological analysis of tissue samples is the cornerstone of cancer diagnosis and treatment but is increasingly strained by rising cancer incidence and a limited number of specialized pathologists~\cite{metter2019Pathologists}.
This growing bottleneck motivates the pressing need for reliable computational methods to assist clinical workflows by screening slides, quantifying morphology, and highlighting diagnostically relevant regions.

Whole Slide Images (WSIs) capture tissue at a gigapixel scale (often exceeding 100,000 × 100,000 pixels) and contain multi-scale morphological cues that are crucial for diagnosis~\cite{Tseng2023HistoGold}.
Modern deep learning pipelines typically handle that scale by partitioning WSIs into small, fixed-size rectangular patches and processing those patches independently~\cite{vanderlaak2021dlinhisto,song2023aiCPath,raciti2023clinicalAIpathImprovement}.
While practical, this patch-based paradigm has two major limitations.
First, rigid patches artificially fragment biological structures (\eg, glands, tumor fronts), destroying natural boundaries and the contextual relationships pathologists use~\cite{ciga2021overcomingPatches}.
Second, many patch-aggregation methods produce predictions that are hard to attribute to clearly delineated regions or interpretable features, which reduces trust and slows clinical adoption.

We address these limitations by introducing a novel framework that transforms WSIs into biologically informed graphs that are interpretable by design.
In contrast to fixed-grid patches, our regions adapt in size to tissue homogeneity: homogeneous areas are merged into larger regions for efficiency, while heterogeneous areas remain finely partitioned to preserve diagnostic detail.

\begin{figure*}[ht]
\includegraphics[width=\textwidth,trim={0 0cm 0 0},clip]{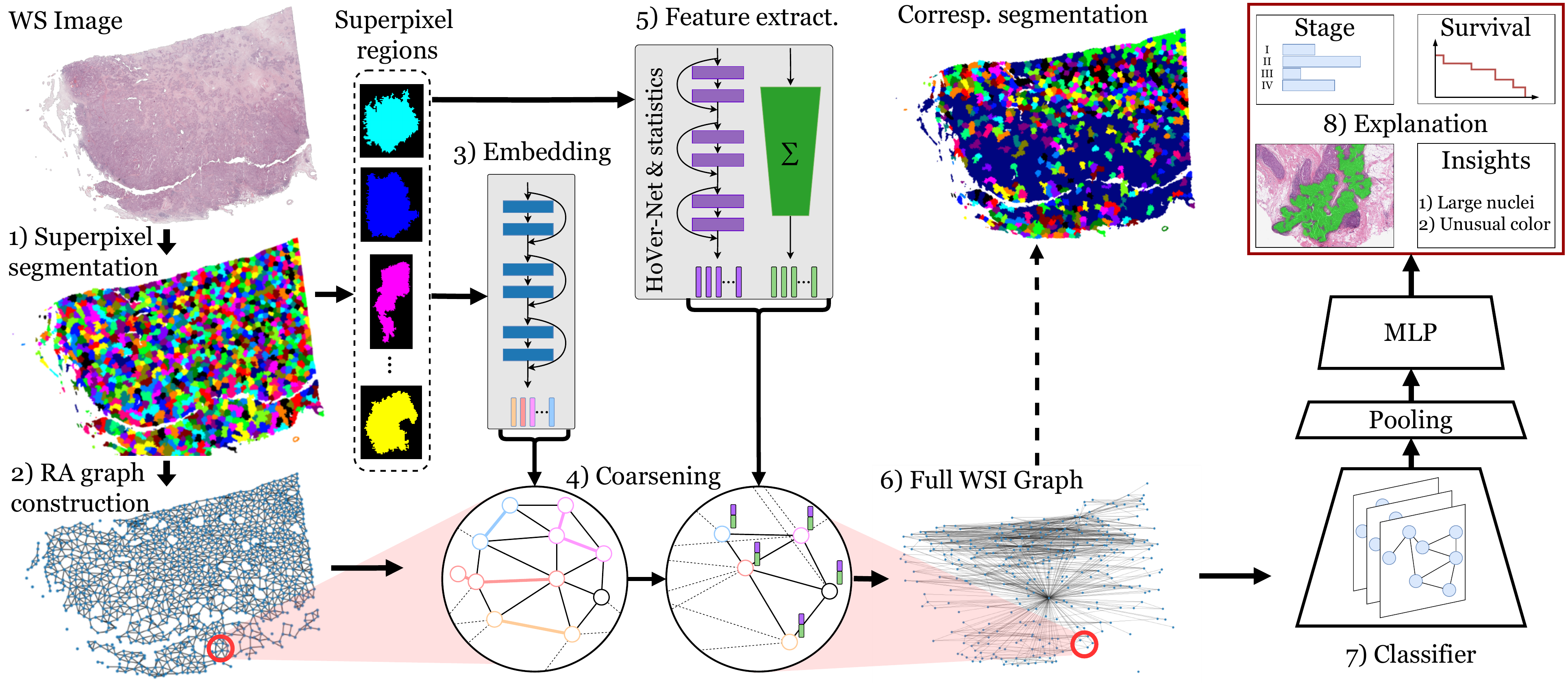}
\caption{Overview of our method for transforming WSIs into graph representations aligning with natural tissue borders.
 Starting from a WSI, superpixels are segmented by clustering pixels based on color and spatial proximity (1) to form a coarse region adjacency (RA) graph $\mathcal{G}_0$ (2). 
 Region embeddings are obtained using a contrastively pretrained CNN (3), enabling similarity-based merging of adjacent nodes (4).
 Interpretable features (texture, morphological, and nuclear characteristics) enrich each node, resulting in a compact yet information-rich representation (6).
 Using this graph, we train a graph attention network for the different diagnostic tasks (7).
 Through integrated gradients, the model's predictions can be attributed to regions and features (8).
}
\label{fig:method}
\end{figure*}

In this paper, we propose a graph-based framework that converts WSIs into biologically meaningful graphs: nodes correspond to tissue regions aligned with natural boundaries, edges capture spatial contiguity, and node descriptors consist of a compact, clinically-interpretable set of texture, morphological, and nuclear features.
Our contributions are: 
\begin{enumerate}
    \item A novel pipeline that builds graphs from adaptive tissue regions, preserving biological context.
    \item An embedding-guided graph coarsening strategy that aggregates homogeneous areas for computational efficiency while retaining high granularity in complex regions.
    \item A demonstration that our compact, interpretable model achieves performance competitive with a large foundation model, while requiring a fraction of the data (300x less), number of parameters (13x less) and computational resources, and providing clinically relevant explanations.
\end{enumerate}
\section{Related work}
\textbf{Multi-Instance Learning (MIL)} is the dominant paradigm for WSI classification~\cite{zaheer2017deepSets,srinidhi2021HistoSurvey,shmatko2022AIinHisto}.
In this framework, each WSI is treated as a bag of thousands of instances (patches), from which a single slide-level label is predicted. 
Seminal models like DeepSets~\cite{zaheer2017deepSets}, DSMIL~\cite{li2021dsmil} and the attention-based ABMIL~\cite{ilse2018abmil} established the effectiveness of this approach.
The performance of patch-based methods has recently culminated in the development of massive foundation models, such as UNI2-h~\cite{chen2024FoundationWSI} and CHIEF~\cite{wang2024CHIEF}.
Pre-trained on hundreds of thousands of slides, these models learn powerful representations of tissue morphology that can be used for whole slide predictions.

These MIL approaches are highly successful for tasks like cancer detection, where the presence of a few key malignant patches can determine the overall diagnosis.
However, for more complex tasks like cancer staging or survival prediction, relying on individual key instances is often insufficient. Such predictions require a more holistic assessment, integrating information from diverse regions.
Furthermore, due to their use of learned patch embeddings, they fail to explain the underlying morphological drivers of the prediction~\cite{kaczmarzyk2024explainable}. This context motivates the exploration of alternative representations, with graph-based methods emerging as a powerful successor capable of explicitly encoding the tissue microenvironment.

\textbf{Graph-based methods} have proven to be powerful tools for analyzing structured data across various domains \cite{wu2021GNNSurvey}. They also offer advantages in spatial structure modeling and interpretability for medical applications~\cite{qiu2024gnnFMri,zheng2022graphDisease,lux2025interpretableretinal,brussee2025gnnHistoSurvey,li2025fine}.
A close extension to standard MIL methods are patch-based graphs.
In those methods, non-overlapping or sliding-window patches serve as graph nodes, and edges are defined by spatial adjacency or k-nearest neighbors.
Architectures like Patch-GCN~\cite{chen2021whole} and the GraphTransformer~\cite{zheng2022GraphTransformer} leverage this structure to allow information to propagate between neighboring patches, capturing local context that is lost in standard MIL.
While this represents an improvement over the bag-of-patches model, it inherits fundamental limitations of its predecessor: the nodes themselves are defined by an artificial grid that fragments natural biological structures, thereby compromising the biological grounding of the graph itself, and a lack of feature interpretability.

Cell-graphs offer the highest-granularity tissue representation by modeling cells as nodes and their interactions as edges~\cite{demir2005augmented,zhou2019cgc,pati2022HACT}. While this provides a high-fidelity map of the cellular microenvironment, it creates a major scalability issue. A single WSI can contain millions of cells, creating graphs that exceed the practical limits of modern GNNs, causing prohibitive memory demands and poor generalization. This limitation restricts cell-graph methods to smaller, expert-annotated regions of interest instead of the entire slide.

Unlike these approaches, our method preserves natural tissue boundaries through adaptive region segmentation. 
Crucially, our coarsening mechanism allows us to create a computationally efficient yet biologically faithful graph representation of the entire WSI.

\section{Method}
Our framework transforms gigapixel-scale WSIs into compact, interpretable graphs, suitable for clinical tasks and explainable predictions.
This transformation addresses the core challenges of computational pathology: the immense scale of the data, the need for both local features and global connections, and the benefit of explainable models in a clinical context.

The proposed pipeline, illustrated in \Cref{fig:method}, constitutes a multi-stage hierarchical process that moves from a low-level pixel representation to a high-level graph suitable for clinical prediction tasks.

The pipeline consists of four main steps. First, the WSI is segmented into a large set of small, biologically aligned regions using superpixels.
Second, these fine-grained regions are adaptively merged based on their semantic similarity, creating a coarsened graph that reflects the macroscopic tissue organization.
Third, each node in this coarsened graph is enriched with a comprehensive set of interpretable, domain-informed features that capture texture, morphology, and nuclear characteristics.
Finally, a Graph Attention Network (GAT) is applied to this final graph representation to perform slide-level classification tasks.

\subsection{Superpixel segmentation}
The first step of our framework is to convert the raw pixel data of a WSI into a meaningful set of initial regions that respect the boundaries of local tissue structures.
More formally, given a WSI $\mathcal{I} \in \mathbb{R}^{H\times W\times 3}$, we first identify the tissue foreground $\mathcal{T} \subset \mathcal{I}$ by applying Otsu's thresholding on the saturation channel of the HSV color channel, followed by morphological operations to remove artifacts, as described in~\cite{lu2021clam}.

\begin{figure}[t]
\includegraphics[width=\linewidth]{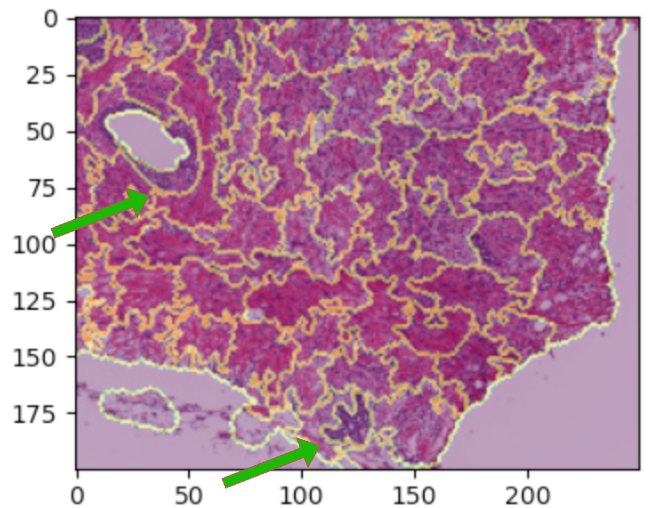}
\caption{Illustration of fine-grained, tissue-adaptive segmentation. Superpixel boundaries, highlighted in yellow, are overlaid on a tissue micrograph to demonstrate their precise alignment with the underlying morphological structures.}
\label{fig:slic}
\end{figure}

With the tissue area $\mathcal{T}$ identified, we then partition it into an initial set of small, perceptually meaningful regions. Instead of using a rigid grid of square patches, we segment the tissue area into superpixels, $\mathcal{S} = \{s_1, ..., s_K\}$, using Simple Linear Iterative Clustering (SLIC)~\cite{radhakrishna2012SLIC} on a low-resolution version of the WSI ($0.625\times$ magnification).
SLIC is a variant of k-means clustering, grouping spatially close pixels with similar colors, creating small, irregularly shaped regions that naturally adhere to local tissue boundaries (see \Cref{fig:slic}), such as gland edges os the interface between tumor and stroma.
The number of superpixels $K$ is chosen to target an average size of $300\times 300$ pixels at a magnification level of x32.
This oversegmentation forms the basis of our initial graph $\mathcal{G}_0 = (\mathcal{V}_0, \mathcal{E}_0)$, where each node $v_i \in \mathcal{V}_0$ corresponds to a superpixel $s_i$, and an edge $(v_i, v_j) \in \mathcal{E}_0$ exists if superpixels $s_i$ and $s_j$ are spatially adjacent.

\subsection{Adaptive graph coarsening}
The initial graph $\mathcal{G}_0$ is too fine-grained for efficient downstream processing.
To address this, we introduce an adaptive graph coarsening procedure designed to merge semantically similar, adjacent regions into larger super-nodes.
The goal is to produce a more abstract, computationally efficient graph, where nodes can represent larger, homogeneous tissue components.
The \textit{adaptive} nature of this process is key: it preserves high granularity in heterogeneous, complex regions while vigorously simplifying large, uniform areas, thus reducing graph size while preserving the most important morphological boundaries.

\noindent\textbf{Region embeddings} The foundation of our coarsening strategy is a semantically rich representation for each graph node.
We generate a region embedding $h_i \in \mathbb{R}^{512}$ for each node $v_i$ using a ResNet-18 feature encoder. The feature encoder was pretrained on histology patches using contrastive learning, enabling it to capture fine-grained textural and morphological patterns \cite{zheng2022GraphTransformer}.

\noindent\textbf{Greedy merging} With these embeddings, we perform a greedy, agglomerative merging of nodes.
Let $A_0$ be the adjacency matrix of the initial graph $\mathcal{G}_0$.
We compute the cosine similarity for all adjacent node pairs $(v_i, v_j)$ in the initial graph $\mathcal{G}_0$:
\begin{equation}
    S_{ij} = \begin{cases}
        \frac{h_i\cdot h_j}{||h_i||\cdot ||h_j||} & \text{if } (A_0)_{ij} = 1 \\
        -\infty & \text{otherwise}
    \end{cases}
\end{equation}

The merge process is performed greedily: starting with the pair with the highest similarity, nodes are merged sequentially until the next candidate pair's score falls below the predefined threshold $\tau$.
The similarity threshold $\tau$ is a hyperparameter that controls the final granularity of the graph.

When two nodes are merged, they are replaced by a single new node whose region is the union of the original nodes, and which inherits the edges of its predecessors.
This agglomerative process continues until no adjacent nodes have a similarity exceeding $\tau$.
The result is a coarsened graph $\mathcal{G} = (\mathcal{V}, \mathcal{E})$ where $|\mathcal{V}|\leq|\mathcal{V}_0|$.

This adaptive coarsening ensures that our final graph is a compact yet faithful representation of the tissue's macro-architecture.

\subsection{Interpretable node features}
While the learned region embeddings $h_i$ used for coarsening are powerful, we do not use them as node features since they are inherently black-box and lack direct clinical interpretability.
To create a framework that is explainable by design, we engineer a separate set of domain-informed features that describe each node in the final coarsened graph.

The feature vector is a concatenation of three distinct feature groups:
$$x_i = x_i^{\text{tex}}\mathbin\Vert x_i^{\text{morph}}\mathbin\Vert x_i^{\text{nuc}}$$

\begin{itemize}
    \item \textbf{Texture and intensity features} $x^{\text{tex}}_i \in \mathbb{R}^{93}$: This group quantifies the micro-patterns within the tissue region. It includes features derived from the Gray-Level Co-occurrence Matrix (GLCM), such as contrast, correlation, and energy, which capture the spatial relationships between pixel intensities. It also includes Local Binary Pattern (LBP) features and general intensity as defined in the Pyradiomics library~\cite{vanGriethuysen2017Radiomics}.
    \item \textbf{Morphological and color features} $x^{\text{morph}}_i \in \mathbb{R}^{21} $: This set of features describes the size of the region and its color properties. We compute color distribution statistics (mean, variance) across multiple color spaces (RGB, HSV, LAB). These features capture tissue morphology and staining properties, which can differentiate between different slides and especially between different hospitals.
    \item \textbf{Nuclear features} $x^{\text{nuc}}_i \in \mathbb{R}^{77} $: Nuclear morphology is a cornerstone of histopathological assessment. We leverage a pretrained HoVerNet model~\cite{graham2019Hovernet}, a frequently-used deep learning model for simultaneous nuclear instance segmentation and classification. For each region, we apply HoVerNet to detect individual nuclei and classify them into one of six types (\eg, neoplastic, inflammatory, stromal)~\cite{gamper2019pannuke}. We then compute a rich set of statistics, including the density, size, and shape characteristics for each nucleus type, providing a quantitative summary of the cellular composition of the region.
\end{itemize}

To reduce redundancy in the high-dimensional initial feature vector, we perform correlation-based pruning. We iteratively remove one feature from any pair in the training data whose absolute Pearson correlation $|\rho|$ exceeds a threshold $\xi$, resulting in a more compact and robust feature set.
A complete list of all features can be found in the supplementary material.

The set of node features $\mathcal{X} = \{x_i \quad | v_i \in \mathcal{V}\}$ extends the previously defined coarsened graph to the compact graph representation $\mathcal{G} = (\mathcal{V}, \mathcal{E}, \mathcal{X})$ of the WSI $\mathcal{I}$.

\subsection{Graph attention network for classification}
Graph Neural Networks (GNNs) are the natural choice for this task, as they explicitly model the relationships between nodes in our graph-structured data. We specifically employ a Graph Attention Network (GAT)~\cite{velickovic2017gat}, which takes our compact representation $\mathcal{G}$ as input and learns the relative importance of neighboring nodes for the final prediction.

\begin{table*}[t]
    \centering
    \setlength\tabcolsep{0pt}
\begin{tabular*}{\textwidth}{@{\extracolsep{\fill}} l ccc c }
        \toprule
        \multirow{2}{*}{\textbf{Method}} & \multicolumn{3}{c}{Stage Classification} & Survival Prediction \\
        \cmidrule(lr){2-4} \cmidrule(lr){5-5}
         & AUC $\uparrow$ & F1$_m$ $\uparrow$ & Balanced Acc $\uparrow$ & C-Index $\uparrow$ \\
        \midrule
        \multicolumn{5}{l}{\textit{Comparable Baselines}} \\
        DeepSets & 52.3$^\ast$ (3.05) & 15.8$^\ast$ (5.50) & 25.4\,\, (0.10) & 48.9$^\ast$ (2.47) \\
        ABMIL    & 61.8$^\ast$ (5.64) & \underline{20.2}$^\ast$ (2.13) & \underline{26.0}\,\, (1.11) & \underline{61.7}\,\, (1.81) \\
        GraphTransformer & \underline{63.3}\,\, (2.33) & 18.6$^\ast$ (0.89) & 24.9\,\, (0.14) & 52.7$^\ast$ (5.61) \\
        Ours & \textbf{67.2}\,\, (3.08) & \textbf{28.0}\,\, (8.24) & \textbf{27.0}\,\, (2.54) & \textbf{62.9}\,\, (3.67) \\
        \midrule
        \multicolumn{5}{l}{\textit{Foundation model}} \\
        UNI2-h & 69.8\,\, (3.05) & 30.1\,\, (4.45) & 31.0$^\ast$(2.78) & 62.1\,\, (6.90) \\
        \bottomrule
\end{tabular*}

    \caption{Performance on TCGA-BRCA (n=1133). Our method is compared against comparable baselines. Best performance among comparable baselines is shown in \textbf{bold}. The foundation model UNI2-h is listed separately for context, as it was pretrained on over 350,000 external WSIs. An asterisk (*) indicates a statistically significant difference ($p < 0.05$) when compared to our method.}

    \label{tab:results-brca}
\end{table*}
\begin{table}[t]
    \centering
     \begin{tabular}{lrrrr}
     \toprule
     & \multicolumn{2}{c}{BRCA} & \multicolumn{2}{c}{UCEC} 
     \\ 
     \cmidrule(lr){2-3} \cmidrule(lr){4-5} 
     Stage & Slides & Ratio & Slides & Ratio
     \\ \midrule 
    I  & 178 & 16.2\%  & 342 & 63.4\%
     \\ 
     II  & 642 & 58.3\%  & 57 & 10.6\%
     \\ 
     III  & 261 & 23.7\%  & 113 & 21.0\%
     \\ 
     IV  & 20 & 1.8\% & 27 & 5.0\%
     \\ 
     \midrule 
     Total  & 1,101 & 100\% & 539 & 100\%
     \\ 
     \bottomrule 
     \end{tabular}

    \caption{Number of patients for each dataset, grouped by their cancer stage label. Both datasets are highly imbalanced in their cancer stage distribution.}
    \label{tab:datasets}
\end{table}

After passing a node through a stack of GAT layers, we obtain a set of high-level node embeddings that are context-aware, incorporating information from their local neighboring tissue environment.
To arrive at a single slide-level prediction, a final graph readout function is applied.
This function aggregates all the node embeddings into a single graph-level feature vector, which is then fed into a standard multilayer perceptron (MLP).
\section{Results \& Discussion}
We evaluate our method on the two challenging clinical tasks of cancer stage classification and patient survival prediction. Our experimental framework, including the datasets, task definitions, and baseline models used for comparison, is detailed below.

\noindent\textbf{Datasets} Our experiments are conducted on two widely used public datasets from The Cancer Genome Atlas (TCGA): the TCGA Breast Invasive Carcinoma (BRCA) dataset~\cite{tcga-brca} and the TCGA Uterine Corpus Endometrial Carcinoma (UCEC) dataset~\cite{tcga-ucec}.
For our evaluation, we use 1,101 hematoxylin and eosin (H\&E) stained WSIs from TCGA-BRCA and 539 from TCGA-UCEC.

\noindent\textbf{Tasks} We benchmark our method on two clinically significant and challenging tasks: cancer stage classification and patient survival prediction.
The first task, cancer stage prediction, involves predicting the pathological stage of the tumor directly from the WSI. Cancer staging is a critical component of clinical oncology, as it describes the extent of cancer progression and acts as a basis to determine treatment plans.
The stage is typically described on a scale from I to IV, with higher stages indicating more advanced disease.
We frame the task according to the common way in literature as a multi-class classification problem where the model must assign one of four stage labels to each slide.

The second task, survival prediction, stratifies patients into risk groups based on their predicted prognosis.
Accurate survival prediction is essential for personalizing treatment, managing patient care, and identifying high-risk individuals who may benefit from more aggressive therapies.
This task is framed as a multi-class classification task, where patients' risks are discretized into four groups, and this assignment acts as the prediction target, following~\cite{wulczyn2020survival}. 

\noindent\textbf{Baselines} We compare our method against two distinct classes of models to ensure a comprehensive evaluation. The first class includes established architectures that are comparable in terms of computational efficiency, while the second class features a state-of-the-art foundation model to benchmark against the upper echelons of current performance.

The first group of baselines consists of representative MIL and graph-based models, which are standard paradigms in computational pathology.
\begin{itemize}
    \item \textbf{DeepSets}~\cite{zaheer2017deepSets} is a permutation-invariant model designed for learning on unordered sets. In the context of WSIs, it treats the collection of image patches as a set, applying a transformation to each patch embedding and then using a summation operation to produce a single feature vector for the entire slide, making it a robust and simple baseline.
    \item \textbf{ABMIL}~\cite{ilse2018abmil} is an attention-based MIL model. It extends the MIL framework by incorporating an attention mechanism that learns to assign different weights to each patch (instance) within the WSI (bag). This allows the model to focus on the most diagnostically relevant regions of the tissue when making a slide-level prediction.
    \item \textbf{GraphTransformer}~\cite{zheng2022GraphTransformer} models the WSI as a graph, where patches are nodes and spatial adjacency defines the edges. This approach explicitly captures the tissue architecture and the spatial relationships between different regions. 
\end{itemize}
\noindent
The second category of comparison involves a recent, large-scale foundation model.

\begin{itemize}
    \item \textbf{UNI2-h}~\cite{chen2024FoundationWSI} serves as our state-of-the-art benchmark. It is a vision foundation model pre-trained on a massive-scale dataset of over 100 million pathology image patches. Such models learn powerful and generalizable representations of tissue morphology that can be usec for downstream tasks as described in~\cite{chen2024FoundationWSI}. Comparing against UNI2-h allows us to assess our method's performance relative to a model that leverages enormous computational resources and data for pre-training, providing critical context for our results.
\end{itemize}
\begin{table*}[th]
    \centering
    \setlength\tabcolsep{0pt}
\begin{tabular*}{\textwidth}{@{\extracolsep{\fill}} l ccc c }
        \toprule
        \multirow{2}{*}{\textbf{Method}} & \multicolumn{3}{c}{Stage Classification} & Survival Prediction \\
        \cmidrule(lr){2-4} \cmidrule(lr){5-5}
         & AUC $\uparrow$ & F1$_m$ $\uparrow$ & Balanced Acc $\uparrow$ & C-Index $\uparrow$ \\
        \midrule
        \multicolumn{5}{l}{\textit{Comparable Baselines}} \\
        DeepSets & 49.4$^\ast$ (1.48) & 9.46$^\ast$ (6.47) & 24.4\,\, (1.67) & 50.3$^\ast$ (0.60) \\
        ABMIL    & 49.0$^\ast$ (2.61) & \underline{18.8}\,\, (0.00) & 25.0\,\, (0.00) & \underline{58.3}\,\, (5.31) \\
        GraphTransformer & \underline{54.1}\,\, (4.51) & 18.7\,\, (1.18) & \underline{25.2}\,\, (0.58) & 57.5\,\, (7.84) \\
        Ours & \textbf{56.4}\,\, (4.17) & \textbf{19.5}\,\, (1.19) & \textbf{26.5}\,\, (2.37) & \textbf{60.0}\,\, (0.97) \\
        \midrule
        \multicolumn{5}{l}{\textit{Foundation Model}} \\
        UNI2-h & 62.7$^\ast$(5.03) & 28.0$^\ast$(4.67) & 30.4$^\ast$(2.92) & 62.8\,\, (5.38) \\
        \bottomrule
\end{tabular*}
    \caption{Performance on TCGA-UCEC (n=566). Our method is compared against comparable baselines. Best performance among comparable baselines is shown in \textbf{bold}. The foundation model UNI2-h is listed separately for context, as it was pretrained on over 350,000 external WSIs. An asterisk (*) indicates a statistically significant difference ($p < 0.05$) when compared to our method.}
    \label{tab:results-ucec}
\end{table*}

\noindent\textbf{Evaluation metrics} We evaluate the performance of models on the cancer staging classification task using AUC, balanced accuracy and the macro-averaged F1-score ($\text{F}1_\text{m}$) to account for the class imbalances (see \Cref{tab:datasets}).
To evaluate models on survival prediction we use the concordance index (c-index)~\cite{harrell1982cindex}.

\noindent\textbf{Experimental setup} To ensure a rigorous and unbiased evaluation, we first partition the entire dataset into a fixed training set and a hold-out test set. All splits are performed at the patient level to prevent data leakage.
Model optimization and selection are conducted solely on the training set. For each method, we perform a 25-trial random search for the learning rate and weight decay. Each trial consists of training five model instances on a sub-partition of the training set and evaluating them on a validation set. The models from the trial yielding the best average validation performance are then selected and their performance on the hold-out test set is reported (mean and standard deviation). To asses statistical significance between our method and the baselines, we apply an independent Student's t-test to the sets of five test scores.

\subsection{Performance analysis with comparable methods}
As shown in \Cref{tab:results-brca} and \Cref{tab:results-ucec}, our graph framework consistently and significantly outperforms all comparable baselines across both datasets and tasks.
To ensure a fair comparison, these baselines are methods trained on the same amount of data, as opposed to large-scale, pre-trained foundation models.

On the larger TCGA-BRCA dataset (\Cref{tab:results-brca}), our method achieves the strongest performance among non-foundation models.
For the complex task of cancer staging, our method achieves an AUC of 67.2 and a macro F1-score of 28.0, a substantial improvement over the next baselines, GraphTransformer (63.3 AUC) and ABMIL (20.2 F1).
The significant boost in the macro-averaged F1-score is particularly noteworthy, as it underscores our model's superior ability to handle the class imbalance in the staging task (see \Cref{tab:datasets} for class distributions).
In survival prediction, our method also achieves the highest c-index (62.9) of all models in this category.

We attribute this performance advantage to our model's fundamentally different approach to representing tissue.
Unlike MIL models that treat a WSI as an unordered bag of artificial patches, or patch-graph methods that impose a grid-based topology, our framework preserves crucial tissue structures.
Specifically, it constructs a graph that mirrors the tissue's macro-architecture by adaptively merging biologically-aligned superpixels, then enriches this structure with a rich suite of interpretable textural, morphological, and nuclear features to capture both microscopic characteristics and their global spatial relationships.

The trend of strong performance continues on the smaller TCGA-UCEC dataset, demonstrating the robustness and good generalizability of our approach.
Once again, our method outperforms the comparable baselines in both cancer staging (AUC of 56.4 vs 54.1, and F1 of 19.5 vs 18.8) and survival prediction (c-index of 60.0 vs 58.3), reinforcing the benefits of our biologically-aligned representations.

\subsection{Comparison to a foundation model}
While foundation models often define the state of the art through scale, our work explores an alternative path: achieving competitive results through intelligent representation design.
Our findings show that this approach can dramatically close the performance gap while offering orders of magnitude gains in training efficiency and improved interpretability.

On TCGA-BRCA, our model's performance is remarkably close to that of UNI2-h.
For survival prediction, our lightweight model slightly surpasses the foundation model, achieving a higher c-index (62.9 vs 62.1).
In the staging task, the performance difference is minimal and, for many metrics, not statistically significant.
This result is remarkable when considering the enormous disparity in resources:
\begin{itemize}
    \item \textbf{Data scale}: UNI2-h was pretrained on a private dataset of over 200 million image patches from 350,000 WSIs. Our model leveraged a pretrained feature encoder (trained on 665 WSIs)~\cite{zheng2022GraphTransformer} and a HoVerNet pretrained on PanNuke (455 visual fields from WSIs)~\cite{gamper2019pannuke}. Counting each visual field as a separate WSI results in 312x less data usage.
    \item \textbf{Model size}: Our final model has 1,003,908 parameters, our region embedding model 11,680,936 parameters and the HoverNet 37,721,166. In total, our pipeline consists of 50,406,010 parameters (1,003,908 trainable). In contrast, UNI2-h has 681,394,176 parameters, with additional 1,183,365 parameters from the stage classification model. In proportion, our pipeline has more than 13x fewer parameters.
    \item \textbf{Feature interpretability}: Beyond efficiency, our model is interpretable by design. Predictions can be traced back to specific, biologically-grounded regions and a curated set of clinically-motivated features. This stands in stark contrast to the black-box features used by most other models and UNI2-h.
\end{itemize}

In essence, our framework presents a compelling alternative for clinical application.
It delivers performance that is competitive with the state-of-the-art, without demanding massive computational infrastructure or sacrificing the ability to understand and validate the model's reasoning.

\subsection{Ablations}
To validate our design choices, we conducted ablation studies on the introduced hyperparameters (\Cref{tab:ablation}).
Our analysis of the feature correlation threshold $\xi$ confirms that a moderate level of feature pruning ($\xi = 0.99$) is beneficial, removing redundant information and improving generalization.
Retaining all features ($\xi = 1.0$) slightly degrades performance, likely due to overfitting.

The study on the region merging similarity threshold $\tau$ highlights another trade-off.
A low $\tau$ (e.g., $0.8$) merges regions more readily, creating a coarser, more computationally efficient graph.
A higher $\tau$ (e.g., $0.95$) preserves more fine-grained detail.
The optimal performance lies within this range, confirming our hypothesis that adaptively simplifying homogeneous areas while preserving detail in heterogeneous ones is key to our model's success.

\subsection{Qualitative analysis and clinical explanations}
\begin{figure*}[ht]
\includegraphics[width=\textwidth]{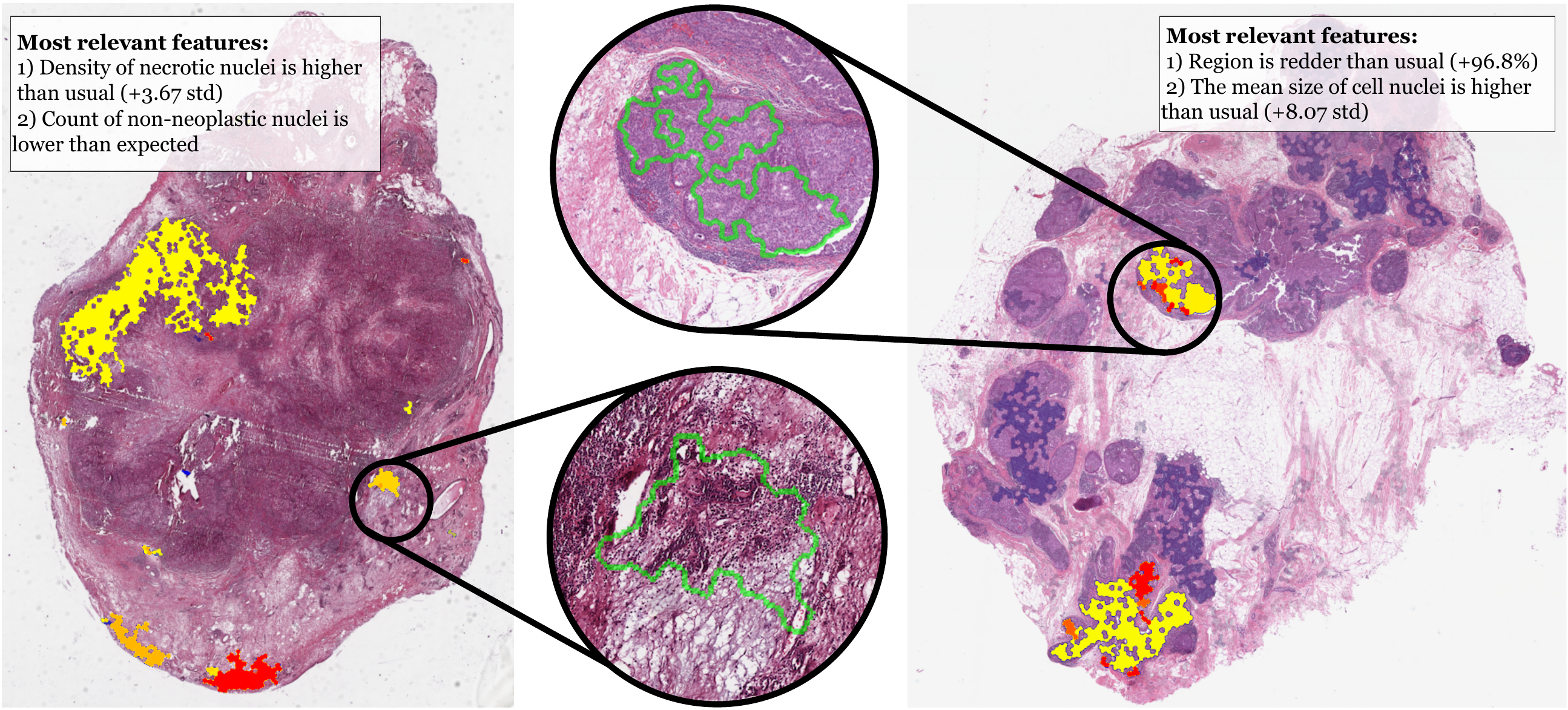}
\caption{Interpretable explanations for two Stage 2 cancer samples from TCGA-BRCA. The overlay highlights the most influential tissue regions (red=high importance). Additionally, we provide feature explanations for both predictions, by naming the most attributed features for the prediction, as well as a comparison to dataset statistics.
}
\label{fig:explainability}
\end{figure*}
\begin{table}[t]
    \centering
    \begin{tabular}{lccc}
        \toprule
        \textbf{Value} & \textbf{AUC} $\uparrow$ & \textbf{F1$_m$} $\uparrow$ & \textbf{Bal. Acc} $\uparrow$ \\
        \midrule
        \multicolumn{4}{c}{\textbf{Correlation Threshold} $\xi$} \\
        0.95  & 62.9 (2.02) & 18.6 (0.25) & 28.1 (3.22) \\%
        0.99  & \textbf{67.4} (2.96) & \textbf{28.4} (2.99) & \textbf{29.6} (1.88) \\%
        1.0   & 61.8 (5.43) & 19.9 (1.30) & 25.5 (0.53) \\%
        \midrule
        \multicolumn{4}{c}{\textbf{Group Similarity} $\tau$} \\
        0.5  & 59.5 (2.29) & 18.6 (0.26) & 25.0 (0.00) \\%
        0.8  & \textbf{64.0} (4.65) & 20.3 (2.81) & 26.6 (1.96) \\%
        0.9  & 63.7 (5.55) & 18.6 (0.26) & 25.0 (0.00) \\%
        0.95 & 62.3 (5.10) & \textbf{25.1} (4.15) & \textbf{28.4} (2.46) \\%
        1.0  & 60.4 (4.98) & 22.1 (2.64) & 25.8 (0.53) \\%
        \bottomrule
    \end{tabular}
    \caption{Ablation study on newly introduced hyperparameters, correlation threshold $\xi$ and group similarity $\tau$ for TCGA-BRCA evaluated on the validation set for cancer stage classification. The best metric per category is \textbf{bold}.}
    \label{tab:ablation}
    
\end{table}

The goal of computational pathology is not just to provide predictions, but to generate actionable insights.
A key motivation for moving beyond patch-based black boxes is the need for trustworthy and clinically relevant explanations.
Our framework delivers on this promise.

By applying Integrated Gradients~\cite{sundararajan2017axiomatic}, we can precisely attribute a prediction to the most influential tissue regions and the specific, interpretable features within them (\Cref{fig:explainability}).
Using the method we can compute an attribution score $\text{IG}_i$ for feature $i$:
\begin{equation}
\text{IG}_i(x) = (x_i - x'_i) \times \sum_{k=1}^{m} \frac{\partial F (x' + \frac{k}{m} \times (x - x'))}{\partial x_i} \times \frac{1}{m} 
\end{equation}
where $x$ is the input, $x'$ is the baseline, $F$ is the model, and $m$ is the number of steps in the approximation.
The baseline $x'$ represents an uninformative input and is constructed using the same graph connectivity but with all node features zeroed out.

The model's interpretability is rooted in its analysis of fundamental tissue components, specifically the different types of cell nuclei it has been trained to distinguish. These biologically relevant classes provide the vocabulary for its explanations:

\begin{itemize}
    \item Neoplastic: These are the tumor cells themselves (malignant or benign), characterized by abnormal growth.
    \item Non-Neoplastic Epithelial: These are normal, hyperplastic, or dysplastic epithelial cells that are not part of the tumor mass.
    \item Inflammatory: Immune system cells, such as lymphocytes and macrophages, responding to the tumor microenvironment.
    \item Connective / Soft Tissue: Cells forming the stroma and supporting tissue, like fibroblasts and endothelial cells.
    \item Dead: Nuclei of cells in apoptotic or necrotic states, often fragmented or degraded, which can be an important biological indicator.
\end{itemize}

Our analysis shows that the most important features for the predictions align with clinical practice, as pathologists routinely use these features for grading~\cite{hayakawa2021nucleiPath,singletary2006BreastStagingGuideline}.
For instance, in \Cref{fig:explainability} (left), the model identifies regions with a high density of necrotic nuclei as drivers for a high-grade prediction.
In \Cref{fig:explainability} (right), it focuses on abnormal nuclear size and unusual color distributions, both key indicators used in practice.
As shown in these examples, we also compare the most attributed features to statistics derived from the training dataset. This can help to provide insights into the reasons for a given prediction and to contextualize the explanation. By understanding whether the prediction is driven by a deviation from common patterns or by the presence of a specific, rare indicator, experts can better validate and build confidence in the model's findings.
For more examples see the supplementary material.

Furthermore, our region-based approach provides a clearer delineation of relevant areas compared to traditional heat maps.

This capability to generate feature-level explanations for specific, biologically coherent regions of interest is a significant step towards building trust with clinicians and provides a powerful tool for validating model behavior.

\section{Conclusion}
We have introduced a novel framework for whole-slide image analysis that represents a significant step towards bridging the gap between high-performance deep learning and the clinical need for interpretability and explainability.
Our core contribution is a shift away from conventional, rigid patch-based methods towards a biologically-grounded, tissue structure adaptive graph representation of WSIs.
Through adaptive segmentation and semantic coarsening, our method captures crucial histological context in a compact and efficient manner, while preserving the essential macro- and micro-structural information of the tissue.

Furthermore, we enrich this structural representation with a multi-faceted feature set that is deliberately designed for interpretability.
By describing each tissue region with a curated set of texture, morphological, and nuclear features, we enable model explanations that can be visually mapped back to the WSI and align with a pathologist's diagnostic process.
This opens the door for a more collaborative human-AI interaction, where the model not only provides a prediction but also highlights the specific morphological evidence supporting its conclusion.

Our experimental results show that our clinically-motivated framework consistently outperforms comparable graph-based and multi-instance learning approaches on two challenging tasks and two datasets.
Furthermore, our method achieves predictive performance that is highly competitive with, and in some cases exceeds, that of the massive state-of-the-art foundation model, UNI2-h.
Crucially, our model accomplishes this without UNI2-h's prohibitive data and computational requirements.
We believe our work shows a promising path for building computer vision systems in digital pathology that are powerful, efficient, and trustworthy.

\textbf{Limitations and future work}
Our proposed framework, while promising, has several limitations that open avenues for future research. The initial graph construction relies on a pre-trained HoVerNet for nuclei detection, a process that is computationally intensive and time-consuming, taking several minutes per whole-slide image. Future work could explore more efficient approaches that potentially accelerate the pipeline.
Future work will focus on large-scale clinical validation of our model's explanations in an expert reader study and exploring methods to directly integrate clinical guidelines into the model's training.

The inherent flexibility of our graph-based framework also enables several exciting extensions.
The integration of multi-modal data, such as genomic or clinical data, is a natural next step. 
Finally, our framework could serve as a powerful tool for scientific discovery. By analyzing the tissue components and morphological features that the model identifies as most important for its predictions, we can potentially uncover novel morphological biomarkers.

{
    \small
    \bibliographystyle{ieeenat_fullname}
    \bibliography{references}
}
\newpage

\onecolumn
\section*{Appendix}

\subsection*{A.1 Additional qualitative examples}

\begin{figure*}[ht]
\includegraphics[width=0.6\linewidth]{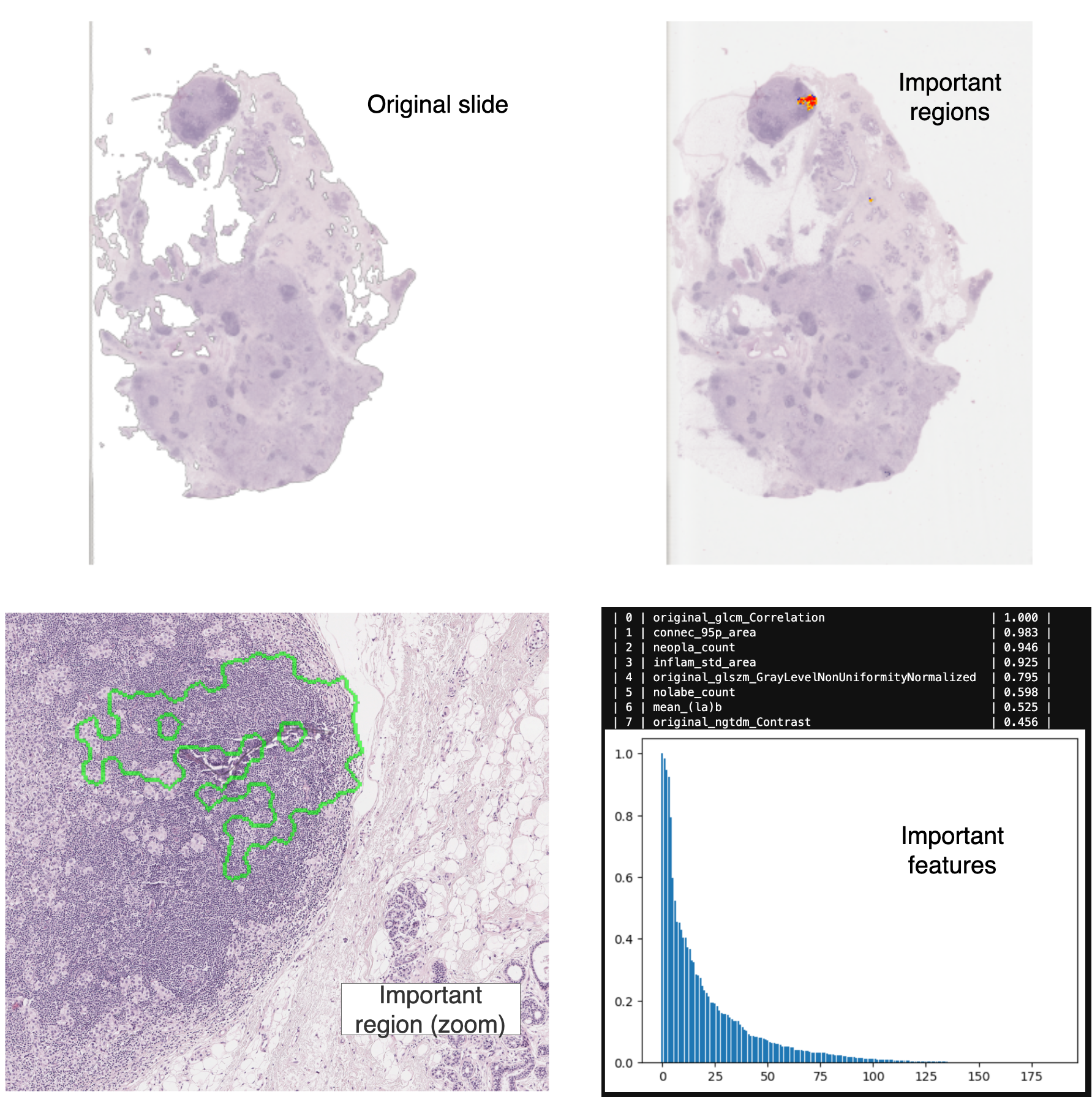}
\centering
\caption{Top left: Original WSI; top right: important regions are highlighted; bottom left: zoomed in version of highlighted region; bottom right: feature attribution scores.}
\end{figure*}

\begin{figure*}[ht]
\includegraphics[width=0.6\linewidth]{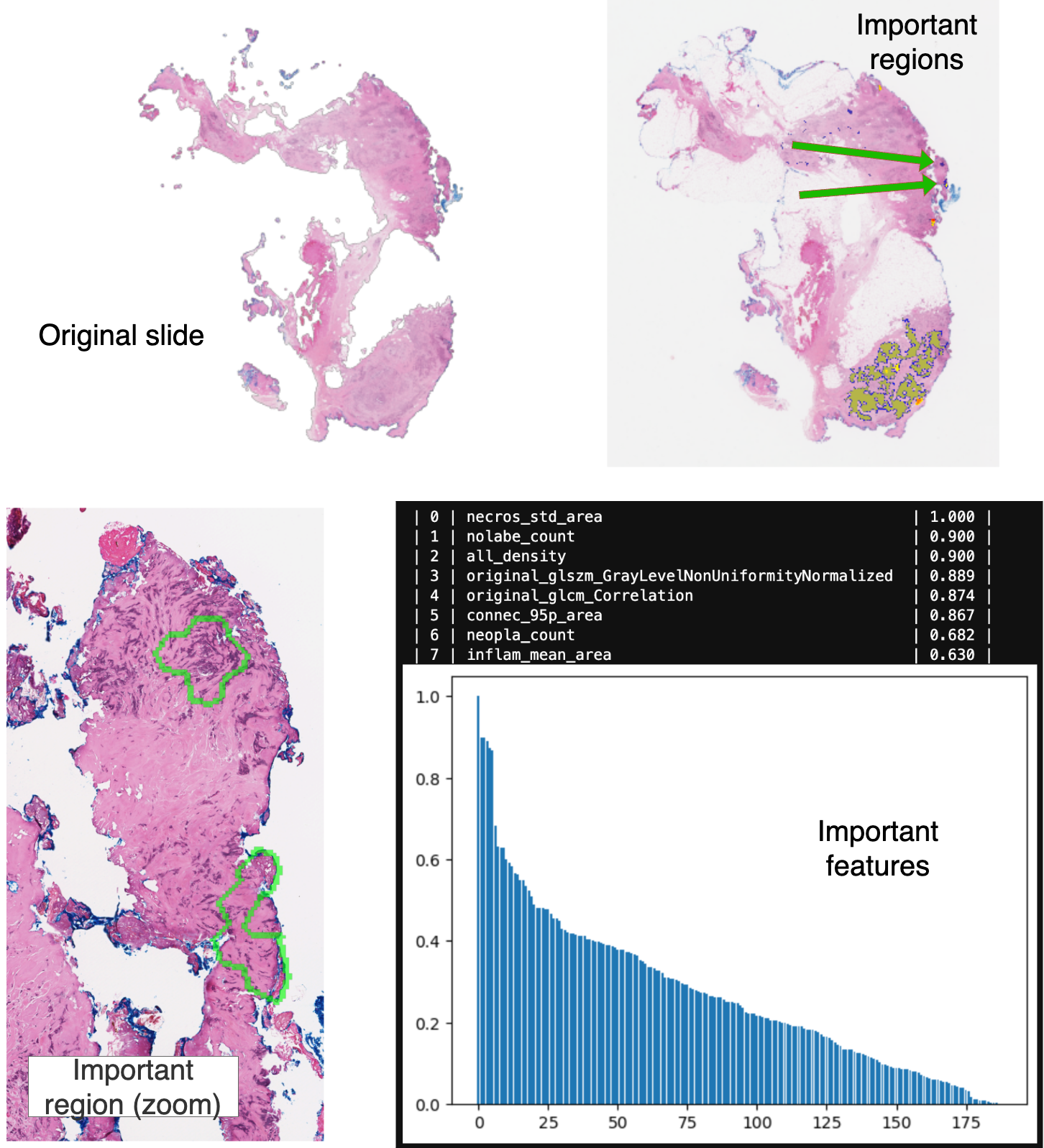}
\centering
\caption{Top left: Original WSI; top right: important regions are highlighted; bottom left: zoomed in version of highlighted region; bottom right: feature attribution scores.}
\end{figure*}

\clearpage

\subsection*{A.2 Feature list}
\footnotesize

\begin{center}
\begin{longtable}{|l|c|c||l|c|c|}
\hline
\textbf{Feature Name} & \textbf{Corr $< 0.95$} & \textbf{Corr $< 0.99$} & \textbf{Feature Name} & \textbf{Corr $< 0.95$} & \textbf{Corr $< 0.99$} \\
\hline
mean\_r & \checkmark & \checkmark & median\_r & \checkmark & \checkmark \\
mean\_g &  & \checkmark & median\_g & \checkmark & \checkmark \\
mean\_b &  & \checkmark & median\_b & \checkmark & \checkmark \\
mean\_h &  & \checkmark & ratio\_bright & \checkmark & \checkmark \\
mean\_s &  & \checkmark & ratio\_dark & \checkmark & \checkmark \\
mean\_v &  & \checkmark & 10\_dark & \checkmark & \checkmark \\
mean\_l &  & \checkmark & 10\_bright & \checkmark & \checkmark \\
mean\_a &  \checkmark & \checkmark & mean & \checkmark & \checkmark \\
mean\_(la)b &  \checkmark & \checkmark & size & \checkmark & \checkmark
\\\hline
\caption{List of morphological and color features $x^{\text{morph}}$ and whether they are included for different correlation thresholds $\tau$.}
\end{longtable}
\end{center}

\footnotesize
\begin{center}
\begin{longtable}{|l|c|c||l|c|c|}
\hline
\textbf{Feature Name} & \textbf{$<0.95$} & \textbf{$<0.99$} & \textbf{Feature Name} & \textbf{$<0.95$} & \textbf{$<0.99$} \\
\hline
\textbf{original\_firstorder} & & & \textbf{original\_glrlm} & & \\
\hspace{2mm}...10Percentile & & \checkmark & \hspace{2mm}...GrayLevelNonUniformity & \checkmark & \checkmark \\
\hspace{2mm}...90Percentile & \checkmark & \checkmark & \makecell{\hspace{2mm}...GrayLevelNonUniformity\\Normalized} & \checkmark & \checkmark \\
\hspace{2mm}...Energy & \checkmark & \checkmark & \hspace{2mm}...GrayLevelVariance & & \checkmark \\
\hspace{2mm}...Entropy & \checkmark & \checkmark & \hspace{2mm}...HighGrayLevelRunEmphasis & & \checkmark \\
\hspace{2mm}...InterquartileRange & \checkmark & \checkmark & \hspace{2mm}...LongRunEmphasis & \checkmark & \checkmark \\
\hspace{2mm}...Kurtosis & \checkmark & \checkmark & \makecell{\hspace{2mm}...LongRunHighGrayLevel\\Emphasis} & \checkmark & \checkmark \\
\hspace{2mm}...Maximum & \checkmark & \checkmark & \makecell{\hspace{2mm}...LongRunLowGrayLevel\\Emphasis} & \checkmark & \checkmark \\
\hspace{2mm}...MeanAbsoluteDeviation & & \checkmark & \hspace{2mm}...LowGrayLevelRunEmphasis & & \checkmark \\
\hspace{2mm}...Mean & & & \hspace{2mm}...RunEntropy & \checkmark & \checkmark \\
\hspace{2mm}...Median & & \checkmark & \hspace{2mm}...RunLengthNonUniformity & \checkmark & \checkmark \\
\hspace{2mm}...Minimum & \checkmark & \checkmark & \makecell{\hspace{2mm}...RunLengthNonUniformity\\Normalized} & \checkmark & \checkmark \\
\hspace{2mm}...Range & & \checkmark & \hspace{2mm}...RunPercentage & & \checkmark \\
\makecell{\hspace{2mm}...RobustMeanAbsolute\\Deviation} & & \checkmark & \hspace{2mm}...RunVariance & \checkmark & \checkmark \\
\hspace{2mm}...RootMeanSquared & & \checkmark & \hspace{2mm}...ShortRunEmphasis & \checkmark & \checkmark \\
\hspace{2mm}...Skewness & \checkmark & \checkmark & \makecell{\hspace{2mm}...ShortRunHighGray\\LevelEmphasis} & \checkmark & \checkmark \\
\hspace{2mm}...TotalEnergy & & & \makecell{\hspace{2mm}...ShortRunLowGray\\LevelEmphasis} & \checkmark & \checkmark \\
\hspace{2mm}...Uniformity & & \checkmark & & & \\
\hspace{2mm}...Variance & \checkmark & \checkmark & & & \\
\hline
\textbf{original\_glcm} & & & \textbf{original\_glszm} & & \\
\hspace{2mm}...Autocorrelation & & \checkmark & \hspace{2mm}...GrayLevelNonUniformity & \checkmark & \checkmark \\
\hspace{2mm}...ClusterProminence & \checkmark & \checkmark & \makecell{\hspace{2mm}...GrayLevelNonUniformity\\Normalized} & \checkmark & \checkmark \\
\hspace{2mm}...ClusterShade & \checkmark & \checkmark & \hspace{2mm}...GrayLevelVariance & \checkmark & \checkmark \\
\hspace{2mm}...ClusterTendency & \checkmark & \checkmark & \makecell{\hspace{2mm}...HighGrayLevelZone\\Emphasis} & \checkmark & \checkmark \\
\hspace{2mm}...Contrast & \checkmark & \checkmark & \hspace{2mm}...LargeAreaEmphasis & \checkmark & \checkmark \\
\hspace{2mm}...Correlation & \checkmark & \checkmark & \makecell{\hspace{2mm}...LargeAreaHighGray\\LevelEmphasis} & \checkmark & \checkmark \\
\hspace{2mm}...DifferenceAverage & \checkmark & \checkmark & \makecell{\hspace{2mm}...LargeAreaLowGray\\LevelEmphasis} & \checkmark & \checkmark \\
\hspace{2mm}...DifferenceEntropy & \checkmark & \checkmark & \makecell{\hspace{2mm}...LowGrayLevelZone\\Emphasis} & \checkmark & \checkmark \\
\hspace{2mm}...DifferenceVariance & \checkmark & \checkmark & \hspace{2mm}...SizeZoneNonUniformity & \checkmark & \checkmark \\
\hspace{2mm}...Id & \checkmark & \checkmark & \makecell{\hspace{2mm}...SizeZoneNonUniformity\\Normalized} & \checkmark & \checkmark \\
\hspace{2mm}...Idm & & & \hspace{2mm}...SmallAreaEmphasis & \checkmark & \checkmark \\
\hspace{2mm}...Idmn & \checkmark & \checkmark & \makecell{\hspace{2mm}...SmallAreaHighGray\\LevelEmphasis} & \checkmark & \checkmark \\
\hspace{2mm}...Idn & \checkmark & \checkmark & \makecell{\hspace{2mm}...SmallAreaLowGray\\LevelEmphasis} & \checkmark & \checkmark \\
\hspace{2mm}...Imc1 & \checkmark & \checkmark & \hspace{2mm}...ZoneEntropy & \checkmark & \checkmark \\
\hspace{2mm}...Imc2 & \checkmark & \checkmark & \hspace{2mm}...ZonePercentage & \checkmark & \checkmark \\
\hspace{2mm}...InverseVariance & \checkmark & \checkmark & \hspace{2mm}...ZoneVariance & \checkmark & \checkmark \\
\hspace{2mm}...JointAverage & & & & & \\
\hspace{2mm}...JointEnergy & & \checkmark & \textbf{original\_ngtdm} & & \\
\hspace{2mm}...JointEntropy & & \checkmark & \hspace{2mm}...Busyness & \checkmark & \checkmark \\
\hspace{2mm}...MCC & \checkmark & \checkmark & \hspace{2mm}...Coarseness & \checkmark & \checkmark \\
\hspace{2mm}...MaximumProbability & \checkmark & \checkmark & \hspace{2mm}...Complexity & \checkmark & \checkmark \\
\hspace{2mm}...SumAverage & \checkmark & \checkmark & \hspace{2mm}...Contrast & \checkmark & \checkmark \\
\hspace{2mm}...SumEntropy & & \checkmark & \hspace{2mm}...Strength & \checkmark & \checkmark \\
\hspace{2mm}...SumSquares & & & & & \\
\hline
\textbf{original\_gldm} & & & & & \\
\hspace{2mm}...DependenceEntropy & \checkmark & \checkmark & & & \\
\hspace{2mm}...DependenceNonUniformity & \checkmark & \checkmark & & & \\
\makecell{\hspace{2mm}...DependenceNonUniformity\\Normalized} & & \checkmark & & & \\
\hspace{2mm}...DependenceVariance & \checkmark & \checkmark & & & \\
\hspace{2mm}...GrayLevelNonUniformity & \checkmark & \checkmark & & & \\
\hspace{2mm}...GrayLevelVariance & & & & & \\
\hspace{2mm}...HighGrayLevelEmphasis & & & & & \\
\hspace{2mm}...LargeDependenceEmphasis & & \checkmark & & & \\
\makecell{\hspace{2mm}...LargeDependenceHighGray\\LevelEmphasis} & \checkmark & \checkmark & & & \\
\makecell{\hspace{2mm}...LargeDependenceLowGray\\LevelEmphasis} & \checkmark & \checkmark & & & \\
\hspace{2mm}...LowGrayLevelEmphasis & \checkmark & \checkmark & & & \\
\hspace{2mm}...SmallDependenceEmphasis & \checkmark & \checkmark & & & \\
\makecell{\hspace{2mm}...SmallDependenceHighGray\\LevelEmphasis} & \checkmark & \checkmark & & & \\
\makecell{\hspace{2mm}...SmallDependenceLowGray\\LevelEmphasis} & \checkmark & \checkmark & & & \\
\hline
\caption{List of texture and intensity features $x^{\text{tex}}$ and whether they are included for different correlation thresholds $\tau$.}
\end{longtable}
\end{center}

\newpage

\footnotesize
\begin{longtable}{|l|c|c||l|c|c|}
\hline
\textbf{Feature Name} & \textbf{$<0.95$} & \textbf{$<0.99$} & \textbf{Feature Name} & \textbf{$<0.95$} & \textbf{$<0.99$} \\
\hline
\textbf{all} & & & \textbf{inflam} & & \\
\hspace{2mm}...count & \checkmark & \checkmark & \hspace{2mm}...count & \checkmark & \checkmark \\
\hspace{2mm}...mean\_area & \checkmark & \checkmark & \hspace{2mm}...mean\_area & \checkmark & \checkmark \\
\hspace{2mm}...std\_area & & \checkmark & \hspace{2mm}...std\_area & \checkmark & \checkmark \\
\hspace{2mm}...5p\_area & \checkmark & \checkmark & \hspace{2mm}...5p\_area & & \\
\hspace{2mm}...25p\_area & \checkmark & \checkmark & \hspace{2mm}...25p\_area & & \checkmark \\
\hspace{2mm}...50p\_area & & \checkmark & \hspace{2mm}...50p\_area & & \\
\hspace{2mm}...75p\_area & & \checkmark & \hspace{2mm}...75p\_area & & \\
\hspace{2mm}...95p\_area & & \checkmark & \hspace{2mm}...95p\_area & & \\
\hspace{2mm}...min\_area & \checkmark & \checkmark & \hspace{2mm}...min\_area & \checkmark & \checkmark \\
\hspace{2mm}...max\_area & \checkmark & \checkmark & \hspace{2mm}...max\_area & & \checkmark \\
\hspace{2mm}...density & \checkmark & \checkmark & \hspace{2mm}...density & \checkmark & \checkmark \\
\hline
\textbf{nolabe} & & & \textbf{connec} & & \\
\hspace{2mm}...count & \checkmark & \checkmark & \hspace{2mm}...count & \checkmark & \checkmark \\
\hspace{2mm}...mean\_area & \checkmark & \checkmark & \hspace{2mm}...mean\_area & \checkmark & \checkmark \\
\hspace{2mm}...std\_area & \checkmark & \checkmark & \hspace{2mm}...std\_area & \checkmark & \checkmark \\
\hspace{2mm}...5p\_area & & & \hspace{2mm}...5p\_area & & \\
\hspace{2mm}...25p\_area & & & \hspace{2mm}...25p\_area & \checkmark & \checkmark \\
\hspace{2mm}...50p\_area & & & \hspace{2mm}...50p\_area & & \checkmark \\
\hspace{2mm}...75p\_area & & & \hspace{2mm}...75p\_area & & \checkmark \\
\hspace{2mm}...95p\_area & & & \hspace{2mm}...95p\_area & & \checkmark \\
\hspace{2mm}...min\_area & & \checkmark & \hspace{2mm}...min\_area & \checkmark & \checkmark \\
\hspace{2mm}...max\_area & & \checkmark & \hspace{2mm}...max\_area & \checkmark & \checkmark \\
\hspace{2mm}...density & \checkmark & \checkmark & \hspace{2mm}...density & \checkmark & \checkmark \\
\hline
\textbf{neopla} & & & \textbf{necros} & & \\
\hspace{2mm}...count & \checkmark & \checkmark & \hspace{2mm}...count & \checkmark & \checkmark \\
\hspace{2mm}...mean\_area & \checkmark & \checkmark & \hspace{2mm}...mean\_area & \checkmark & \checkmark \\
\hspace{2mm}...std\_area & \checkmark & \checkmark & \hspace{2mm}...std\_area & \checkmark & \checkmark \\
\hspace{2mm}...5p\_area & & \checkmark & \hspace{2mm}...5p\_area & & \\
\hspace{2mm}...25p\_area & & \checkmark & \hspace{2mm}...25p\_area & & \checkmark \\
\hspace{2mm}...50p\_area & & & \hspace{2mm}...50p\_area & & \\
\hspace{2mm}...75p\_area & & & \hspace{2mm}...75p\_area & & \\
\hspace{2mm}...95p\_area & & \checkmark & \hspace{2mm}...95p\_area & & \\
\hspace{2mm}...min\_area & \checkmark & \checkmark & \hspace{2mm}...min\_area & \checkmark & \checkmark \\
\hspace{2mm}...max\_area & \checkmark & \checkmark & \hspace{2mm}...max\_area & \checkmark & \checkmark \\
\hspace{2mm}...density & \checkmark & \checkmark & \hspace{2mm}...density & \checkmark & \checkmark \\
\hline
\multicolumn{3}{|l||}{} & \textbf{no-neo} & & \\
\multicolumn{3}{|l||}{} & \hspace{2mm}...count & \checkmark & \checkmark \\
\multicolumn{3}{|l||}{} & \hspace{2mm}...mean\_area & \checkmark & \checkmark \\
\multicolumn{3}{|l||}{} & \hspace{2mm}...std\_area & \checkmark & \checkmark \\
\multicolumn{3}{|l||}{} & \hspace{2mm}...5p\_area & & \\
\multicolumn{3}{|l||}{} & \hspace{2mm}...25p\_area & & \checkmark \\
\multicolumn{3}{|l||}{} & \hspace{2mm}...50p\_area & & \\
\multicolumn{3}{|l||}{} & \hspace{2mm}...75p\_area & & \\
\multicolumn{3}{|l||}{} & \hspace{2mm}...95p\_area & & \\
\multicolumn{3}{|l||}{} & \hspace{2mm}...min\_area & \checkmark & \checkmark \\
\multicolumn{3}{|l||}{} & \hspace{2mm}...max\_area & & \checkmark \\
\multicolumn{3}{|l||}{} & \hspace{2mm}...density & \checkmark & \checkmark \\
\hline
\caption{List of nuclear characteristics $x^{\text{nuc}}$ and whether they are included for different correlation thresholds $\tau$.}
\label{tab:feature_nuclei}
\end{longtable}

\end{document}